\documentclass [a4paper,twocolumn]{article}

\usepackage[utf8]{inputenc}
\usepackage[english]{babel}

\usepackage{geometry}
\usepackage{graphicx}
\usepackage{physics}
\usepackage{braket}
\usepackage{bm}
\usepackage{float}
\usepackage{siunitx}
\usepackage{booktabs}

\usepackage{url}
\usepackage[affil-it]{authblk}
\usepackage{blindtext}
\usepackage{abstract}
\usepackage{comment}
\usepackage{amsmath}
\usepackage{amssymb}
\usepackage{bbold}

\DeclareUnicodeCharacter{03B1}{$\alpha$}
\DeclareUnicodeCharacter{03B3}{$\gamma$}

\usepackage{xcolor}
\usepackage{adjustbox}

\usepackage{caption}

\title{Mixed-Valent Magnetism in CeFe$_2$ from Multi-Impurity DFT+DMFT}
\author[1]{B. Herzog}
\author[1,2]{V. Borisov}
\author[1,2]{O. Eriksson$^{*}$}
\affil[1]{Department of Physics and Astronomy, Uppsala University, Box 516, 751 20 Uppsala, Sweden}
\affil[2]{Wallenberg Initiative Materials Science (WISE), Uppsala University}

\date{}

\begin{document}
\twocolumn[
  \begin{@twocolumnfalse} 
    
    \maketitle 

    * Corresponding author, email: olle.eriksson@physics.uu.se

\begin{abstract}
The microscopic origin of magnetism in CeFe$_2$ has remained unresolved for almost forty years, where polarized-neutron diffraction and 
x-ray magnetic circular dichroism infer markedly different Ce $4f$ spin and orbital moments, that also are in disagreement with theory.
We show here that 
within 
a relativistic multi-impurity DFT+DMFT framework, where the 
Fe $3d$ states are treated by spin-polarized T-matrix fluctuation exchange and Ce $4f$ orbitals by a bath-coupled configuration-interaction solver, this long standing problem is resolved. This level of theory is exclusive in reproducing magnetic moments (spin and orbital) for both the Ce and Fe atoms, yielding a total moment in agreement with the measured saturation moment. The theory put forth here is much closer to the atom specific moments reported from XMCD, compared to values from polarized-neutron diffraction.
The occupation $\langle n_f\rangle=0.85$ and charge variance $\delta n_f^2=0.27$ establish substantial valence fluctuations, 
while the spectral function simultaneously recovers significant weight at the Fermi level together with separate incoherent structures. 
These results identify bath-mediated polarization and configuration mixing as the essential ingredients governing the electronic structure and magnetism of 
CeFe$_2$.
\vspace{1cm}
\end{abstract}
\end{@twocolumnfalse}
]

\section{Introduction}
CeFe$_2$ is a cubic C15 Laves-phase ferromagnet with a Curie temperature of approximately $230$~K and a low-temperature saturation moment near $2.3\,\mu_B$ per formula unit \cite{Konishi2000}.
This material is unusual among rare-earth-iron Laves phases because the Ce $4f$ states lie close to the Fermi energy and hybridize strongly with the Fe $3d$ bands. 
This hybridization contributes to the chemical bonding and causes a  contracted lattice compared to that of purely localized, trivalent RE$_2$ Laves phases (here RE can e.g. be Gd or Lu)\cite{Eriksson1988}. The magnetic properties are also anomalous, where a parallel alignment of the Fe and Ce moment, that would be expected for a localized, trivalent light-rare-earth element in contact with Fe\cite{szpunzar}, is replaced with an antiparallel coupling. In addition, experiments observe a significantly reduced Ce moment compared to that of a localized, trivalent atom\cite{Eriksson1988,Konishi2000}. 
Photoemission, inverse photoemission, and x-ray absorption also exclude a description of a purely trivalent, localized Ce ion (or a passive tetravalent ion) in favor of a mixed-valent state\cite{Konishi2000}.

Early spin-polarized band calculations, based on the local spin density approximation (LSDA) to density functional theory (DFT) reproduced with rather good accuracy the total magnetic moment per unit cell of the material, in addition to the antiparallel alignment between Ce and Fe moments, as well as the contracted lattice parameter. 
The calculated magnetization density was strongly nonspherical and its division into atomic and orbital contributions depended on the spatial partition \cite{Eriksson1988,Trygg1994}. Most disturbingly, it was found that even-though the total moment of single determinant theory (such as LSDA-DFT) came out in decent agreement with observations, the individual moments of the Fe and Ce atoms deviates quite dramatically from the experimental values. 
Polarized-neutron diffraction assigned spin and orbital moments of about $-0.10\,\mu_B$ and $+0.03\,\mu_B$ for Ce, whereas x-ray magnetic circular dichroism (XMCD) of the same moments 
obtained values of $-0.37\,\mu_B$ and $+0.21\,\mu_B$ \cite{Kennedy1993,Delobbe1998}. Subsequent XMCD measurements showed that this contrast is not equally robust for the two components\cite{Saitoh2017}. Extraction of the Ce $4f$ spin moment depends strongly on the $jj$ mixing correction and the magnetic dipole term $T_z$, whereas the orbital sum rule is comparatively insensitive to these assumptions. Furthermore, the more approximate theory (using the atomic sphere approximation) of Refs.\cite{Eriksson1988} reports an Fe moment of $+1.48\,\mu_B$ and a Ce moment of $-0.57\,\mu_B$, while the more accurate full potential (LSDA-DFT) calculation of Ref.\cite{Trygg1994} reports an Fe moment of $+1.82\,\mu_B$ and a Ce moment of $-0.56\,\mu_B$. 
The latter calculation relies on a full potential treatment, and it is rather troublesome that it results in a total moment close to $3.0\,\mu_B$ per formula unit, which deviates markedly from the measured value of $2.3\,\mu_B$ per formula unit \cite{Konishi2000}.

CeFe$_2$ is hence a unique system in that none of the probes that is used to investigate it is consistent with any other probe, not when comparing theory to experiment nor when comparing different experimental techniques. However, one must conclude that the measured saturation moment of CeFe$_2$ is robust. This measurement does not rely on complications in interpretation of measured data, in contrast to the XMCD and the neutron measurements. A reliable theory should therefore be able to reproduce the measured saturation moment preferably within 5-10 \%.
Despite the complications in understanding the microscopic nature of the magnetism of CeFe$_2$ this system has been discussed for applications, for instance as a refrigerant material in magneto-caloric applications\cite{magcal} and as a material that provides improved permanent magnets\cite{permmag} for green energy applications and electrification of our society.

Due to the presence of occupied 4f states of the Ce atom in CeFe$_2$, it is likely that the functionals used in previous theory \cite{Eriksson1988,Trygg1994} were inaccurate in capturing correlation effects, and that one should move beyond a single Slater determinant description of this system. One way to estimate the balance of kinematic effects (that is accurately described by conventional - LSDA based -electronic structure theory) and local correlations (that require a multi-configurational description) is to investigate the hybridization function, a property that describes the tendency of electrons to become itinerant. For Ce systems, this was investigated in Ref.\cite{heike} and it was concluded that CeFe$_2$ (together with CeCo$_2$, CeNi$_2$, CeRu$_2$, CeIr$_2$ and many other Ce based Laves phases) have a ballance between kinematic and local Coulomb effects that is similar to that of the $\alpha$-phase of Ce. This system is a well known correlated electronic structure system and the most successful theoretical descriptions of it have been based on dynamical mean field theory (DMFT). Examples of publications using DMFT applied to the $\alpha$-phase of Ce can be found in Refs.\cite{ce_alpha_1,ce_alpha_2,ce_alpha_3,Herzog2025}.

In this paper we have applied DMFT to elucidate the influence of correlations and multi-configurational effects on the electronic structure and the magnetic properties of CeFe$_2$, to try to address an unresolved problem that has haunted the magnetism community for close to forty years. 
Calculations based on DMFT allow to study atomic multiplet formation, mixed valency, and itinerancy on equal footing, something which poses an enormous challenge for conventional electronic-structure theory. There are several levels of approximations to DMFT, in particular how the so called impurity problem is handled. Here we compare the Hubbard-I approximation, the spin-polarized T-matrix fluctuation exchange (SPTF) method, and a configuration interaction approach, that are all described below. 

\section{Computational methods}

We performed relativistic DFT+DMFT calculations using the full-potential LMTO code RSPt \cite{Granas2012}, starting from a non spin-polarized LDA Hamiltonian. The Fe $3d$, Ce $5d$ and Ce $4f$ shells were treated as separate correlated subspaces using the spin-polarized T-matrix fluctuation exchange (SPTF) \cite{Katsnelson2002}, static Hartree-Fock, and either Hubbard-I\cite{HIAsolver} or CLIC solver\cite{Zgid2012,Herzog2025}, respectively. 
The interaction parameters were $(U,J) = (3.0,0.75),(0.10,0.01)$ and $(6.0,0.60)$ eV for Fe $3d$, Ce $5d$, and Ce $4f$, at $k_BT = 0.001$ Ry. Because Hubbard-I removes the hybridization bath, its charge self-consistent calculation included an auxiliary $f-d$ exchange field. 
This term was absent in CLIC, whose one-shot self-energy was inserted into the converged SPTF+HIA electronic structure. The Ce $5d$ shell was retained as a separate, weakly interacting subspace to resolve their induced itinerant polarization and their contribution to the spin-polarized environment of the Ce $4f$ shell. The small interaction parameters produce only a weak static correction, leaving the Ce $5d$ electronic structure essentially LDA-like. By starting from a non-spin-polarized LDA Hamiltonian, we ensure that the ordered state and the Ce polarization arise entirely from the many-body self-energies. A full description of the calculations are provided in Appendix A. 

All DMFT projectors and all reported DMFT spin and orbital moments were defined using the muffin-tin heads. 
Within the present setup the interstitial region carries no magnetic degree of freedom. 

For reference, we also report a conventional spin-polarized LSDA calculation. 
Its local entries use muffin-tin-resolved quantities, whereas its formula-unit value is obtained from the full cell and includes the interstitial contribution. 
It is thus methodologically distinct from the nonmagnetic LDA starting point of the DMFT calculation. 

\section{Results}

The calculated and measured moments are compared in Fig.~\ref{fig:moments} and Table~\ref{tab:moments} of the appendix (Appendix B). 
The spin-polarized LSDA reference gives the correct antiparallel Ce-Fe alignment and a full-cell moment of approximately $2.70\,\mu_B$ per formula unit. This value differs with $\sim$ 10 \% from the previous LSDA value of Ref.\cite{Trygg1994}, a difference that is due to inclusion of 3s and 3p Fe states in the variational step of the Kohn-Sham equation, and a denser k-point sampling in the present calculation\cite{LSDAcheck}.
On this level of theory, and in agreement with the calculation of Ref.\cite{Trygg1994}, the local Fe and Ce components are significantly larger than the experimental values (see Table~\ref{tab:moments}).

\begin{figure*}[t]
  \centering
  \includegraphics[width=0.98\textwidth]{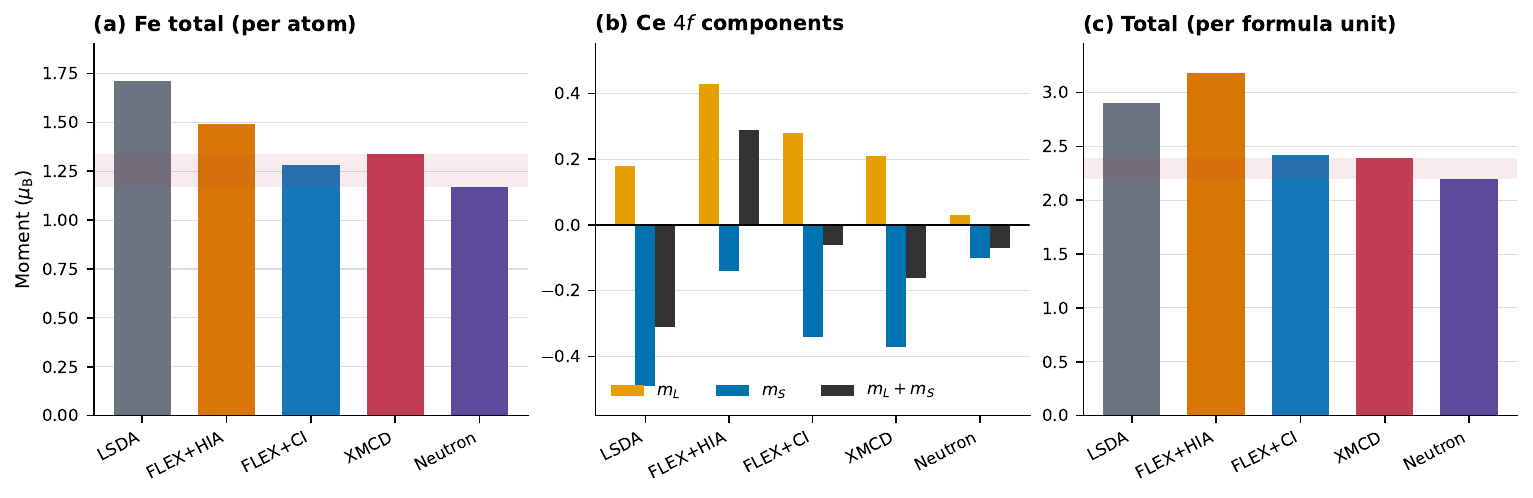}
  \caption{Calculated and experimental magnetic moments. (a) Total Fe moment per atom. (b) Ce-$4f$ orbital, spin, and net moments. (c) Formula-unit values. Positive moments are parallel to Fe. The theoretical local moments are projected atomic centered, muffin-tin quantities. The total moment from the LSDA calculation is a full-cell result including the interstitial contribution, whereas the SPTF+HIA and SPTF+CLIC values are the complete moments of the MT-head magnetic representation. The shaded interval in panel (c) spans the two quoted experimental totals.}
  \label{fig:moments}
\end{figure*}

The DMFT calculation that uses SPTF for the Fe 3d states and HIA for Ce 4f states, reduces the Fe moment relative to the LSDA reference, but gives an orbital-dominated Ce-$4f$ state. 
We obtain for Ce that $m_L^{4f}=+0.43\,\mu_B$, which exceeds the magnitude of $m_S^{4f}=-0.14\,\mu_B$, leaving a net $4f$ contribution of $+0.29\,\mu_B$ parallel to Fe. 
This coupling disagrees with all experimental decompositions. The corresponding formula-unit value of the total moment is $3.18\,\mu_B$ within the MT-head representation, which exceeds the experimental value with a large amount. 

The bath-coupled CLIC solver of the Ce site, combined with SPTF on the Fe site, changes the results both as regards the magnitude and balance of the Ce components. This calculation gives $m_S^{4f}=-0.34\,\mu_B$ and $m_L^{4f}=+0.28\,\mu_B$,
and thus a total Ce 4f moment $m^{4f}=-0.06\,\mu_B$. 
This calculation hence gives total-, spin- and orbital moments in good agreement with the XMCD results of Refs.\cite{Kennedy1993,Delobbe1998}, see Fig.~\ref{fig:moments} and Table~\ref{tab:moments}. In addition, the 5d moment of this calculation ($-0.08\,\mu_B$) is in decent agreement with the measured value.
The Fe moment becomes $1.28\,\mu_B$ per atom, a value that is significantly reduced from the value of the LSDA calculation. In comparison to observations, we note that the calculations based on CLIC and SPTF gives an Fe moment that lies between those given by the XMCD and neutron measurements.
Together these contributions produce a total moment of $2.42\,\mu_B$ per formula unit, in good agreement with the measured saturation moment of 
Ref.\cite{Konishi2000}.

The CLIC + SPTF solution has a Ce-$4f$ occupation of $n_f = 0.85$, which is in good agreement with the conclusion of the resonant photoemission study of Ref. \cite{Jung2001}. The charge variance is $\delta n^2_f = 0.27$, with a corresponding root-mean-square fluctuation of $\sqrt{\delta n_f^2}=0.52$ electrons. These values show directly that the Ce atom, as described by CLIC, is far from a pure $4f^1$ atomic state and enters a mixed valence regime.


\begin{figure}[t]
  \centering
  \includegraphics[width=\columnwidth]{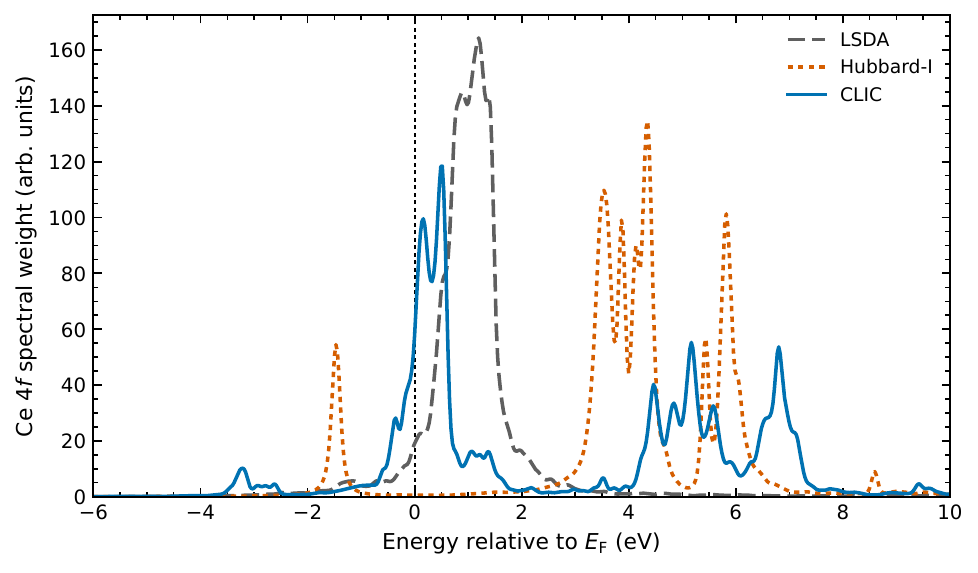}
  \caption{Calculated Ce-$4f$ one-particle spectra for the spin-polarized LSDA reference, SPTF+HIA, and SPTF+CLIC. All curves are shown on their native energy scale and with the same $100$~meV Gaussian broadening. No empirical energy shift, separate intensity normalization, photoemission matrix element, Fermi factor, or background has been applied. The vertical line marks the Fermi energy.}
  \label{fig:ce4fspectra}
\end{figure}

Figure~\ref{fig:ce4fspectra} provides an independent validation of the many-body state underlying the improved moment distribution.
The spin-polarized LSDA reference calculation places most of the Ce-$4f$ spectral weight in a comparatively narrow unoccupied manifold, with a dominant maximum near $+1$~eV. In this calculation, there is  
no separate many-body removal of spectral weight, nor any satelites or other additional structures, in disagreement with experiment\cite{Jung2001}. 
The Hubbard-I approximation produces atomic-like occupied and unoccupied multiplets, including a removal feature near $-2$~eV, 
while leaving essentially no spectral weight at $E_F$, which does not agree with observations. 
Calculations based on CLIC combined with SPTF finds a finite and narrow low-energy spectral weight at the Fermi level, with a broad occupied satelite structure centered near $-3$~eV, and a multiplet-like addition spectrum in the unoccupied states, that is extending 
over several electron volts. Among the three approximations, this level of theory is therefore unique in displaying both the low binding-energy hybridization feature 
and separated incoherent weight expected for a correlated mixed-valent shell. Overall, this calculated spectrum is consistent with measurements. 



\section{Discussion and Conclusion}

The HIA-CLIC comparison changes the physical description in two related ways. 
First, CLIC introduces bath-induced charge and configuration fluctuations, quantified above by $\langle n_f\rangle$ and $\delta n_f^2$. 
Second, the spin-dependent hybridization itself transmits the magnetic environment of Fe and Ce $5d$ to the Ce $4f$ shell. 
Hubbard-I lacks this channel and requires a separate model exchange field. 
CLIC removes that phenomenological term and nevertheless changes the net Ce $4f$ contribution from parallel to antiparallel alignment. 
The central result obtained here is therefore not solely a better numerical value of atomic projected-, or total-, magnetic moments; it is the emergence of the correct polarization mechanism in a mixed-valent many-body state.

The CLIC solution provides a common microscopic framework for the apparently different XMCD and neutron decompositions. 
These techniques probe different projections of the magnetization: XMCD applies element- and shell-specific sum rules, 
whereas neutron diffraction extracts moments through finite-wave-vector magnetic form factors. 
Within these distinct experimental projections, CLIC reproduces essentially all of the magnetic information obtained by the XMCD measurements, as well as the total Ce moment of the neutron scattering experiments and 
the measured total saturation magnetization moment. 
This cross-observable consistency is a central strength of the bath-coupled solution to the DMFT approach. This solution also adds to the discussion on the distinct difference of the moments obtained from XMCD and neutron scattering investigations, and we conclude that an element specific, multi-configuration approach that accurately treats kinematic and local Coulomb effects, is in favor of the XMCD results.

The sharper test is whether a theory obtains a physically consistent Fe-Ce decomposition and represents the experimentally established mixed valence. 
The success of the LSDA formula-unit moment demonstrates that the total magnetization alone is not a sufficiently discriminating benchmark. 
CLIC goes beyond this static agreement by simultaneously reproducing the Fe--Ce moment balance, the antiparallel Ce contribution, 
substantial valence fluctuations, and the characteristic organization of the Ce $4f$ spectrum. 
This simultaneous agreement across ground-state and one-particle observables is the decisive improvement over both LSDA and Hubbard-I approximations.

Taken together, our results show 
converging evidence that the explicit Ce hybridization bath is 
an essential ingredient for describing the electronic structure and magnetism of CeFe$_2$, when combined with local Coulomb interaction (the Hubbard U). 
This is resolved here, by the CLIC solver, that also removes the need for a phenomenological $f-d$ exchange field. When combined with a crucially important description of correlations also on the Fe 3d orbitals, one can arrive to a description of magnetism that is in acceptable agreement with measurements and shows strong valence fluctuations of the Ce 4f states. We end by noting that similar, unresolved aspects of magnetic compounds with 3d elements and f-electron systems, also crystallizing in the Laves structure, are known. CeFe$_2$ is not an isolated case, but the first one addressed here on a DMFT level of electronic structure theory. 

\section{Acknowledgements}

O.E. acknowledges support from the Wallenberg Initiative
Materials Science for Sustainability (WISE) funded by the Knut and Alice Wallenberg Foundation (KAW) and the European Research Council through the ERC Synergy Grant 854843-FASTCORR. O.E. also acknowledges support from STandUPP, eSSENCE, the Swedish Research Council
(VR) and the Knut and Alice Wallenberg Foundation (KAW- Scholar program)
and NL-ECO: Netherlands Initiative for Energy-Efficient Computing (with
project number NWA. 1389.20.140) of the NWA research program. B.H. acknowledges valuable discussions with Patrik Thunström. V.B. acknowledges support by the Swedish Research Council through grant number 2024.05206. The computations were enabled by resources provided by the National Academic Infrastructure for Supercomputing in Sweden (NAISS) at the National Supercomputer Centre (NSC), Linköping University, on the Tetralith supercomputer.

\bibliographystyle{unsrt}
\bibliography{bibli}

@article{Eriksson1988,
  author  = {Eriksson, O. and Nordstrom, L. and Brooks, M. S. S. and Johansson, B.},
  title   = {$4f$-band magnetism in {CeFe$_2$}},
  journal = {Physical Review Letters},
  volume  = {60},
  pages   = {2523--2526},
  year    = {1988},
  url ={http://dx.doi.org/10.1103/PhysRevLett.60.2523},
  doi     = {10.1103/PhysRevLett.60.2523}
}

@article{Kennedy1993,
  author  = {Kennedy, S. J. and Brown, P. J. and Coles, B. R.},
  url = {http://dx.doi.org/10.1088/0953-8984/5/29/012},
  doi = {10.1088/0953-8984/5/29/012},
  title   = {A polarized neutron study of the magnetic form factors in {CeFe$_2$}},
  journal = {Journal of Physics: Condensed Matter},
  volume  = {5},
  number  = {29},
  pages   = {5169--5178},
  year    = {1993}
}

@article{Trygg1994,
  author  = {Trygg, J. and Wills, J. M. and Johansson, B. and Eriksson, O.},
  title   = {First-principles study of the magnetization density in {CeFe$_2$}},
  journal = {Physical Review B},
  volume  = {50},
  pages   = {4200--4203},
  year    = {1994},
  url = {http://dx.doi.org/10.1103/PhysRevB.50.4200},
  doi     = {10.1103/PhysRevB.50.4200}
}

@article{Delobbe1998,
  author  = {Delobbe, A. and Dias, A.-M. and Finazzi, M. and Stichauer, L. and Kappler, J.-P. and Krill, G.},
  title   = {X-ray magnetic circular dichroism study on {CeFe$_2$}},
  journal = {Europhysics Letters},
  volume  = {43},
  number  = {3},
  pages   = {320--325},
  year    = {1998},
  url = {http://dx.doi.org/10.1209/epl/i1998-00359-2},
  doi     = {10.1209/epl/i1998-00359-2}
}

@article{Konishi2000,
  author  = {Konishi, T. and Morikawa, K. and Kobayashi, K. and Mizokawa, T. and Fujimori, A. and Mamiya, K. and Iga, F. and Kawanaka, H. and Nishihara, Y. and Delin, A. and Eriksson, O.},
  title   = {Electronic structure of the strongly hybridized ferromagnet {CeFe$_2$}},
  journal = {Physical Review B},
  volume  = {62},
  pages   = {14304--14312},
  year    = {2000},
  url     = {http://dx.doi.org/10.1103/PhysRevB.62.14304},
  doi     = {10.1103/PhysRevB.62.14304}
}

@article{Jung2001,
  author  = {Jung, R.-J. and Kim, H.-D. and Choi, B.-H. and Oh, S.-J. and Cho, E.-J. and Iwasaki, T. and Sekiyama, A. and Imada, S. and Suga, S. and Park, J.-G.},
  title   = {Temperature-dependent bulk-sensitive {Ce} $3d$ edge resonant photoemission study of {CeFe$_2$}},
  journal = {Journal of Electron Spectroscopy and Related Phenomena},
  volume  = {114--116},
  pages   = {693--698},
  year    = {2001},
  url     = {http://dx.doi.org/10.1016/S0368-2048(00)00360-1},
  doi     = {10.1016/S0368-2048(00)00360-1}
}

@article{Saitoh2017,
  author  = {Saitoh, Y. and Yasui, A. and Fuchimoto, H. and Nakatani, Y. and Fujiwara, H. and Imada, S. and Narumi, Y. and Kindo, K. and Takahashi, M. and Ebihara, T. and Sekiyama, A.},
  title   = {Experimental observation of temperature and magnetic-field evolution of the $4f$ states in {CeFe$_2$} revealed by soft x-ray magnetic circular dichroism},
  journal = {Physical Review B},
  volume  = {96},
  pages   = {035151},
  year    = {2017},
  url = {http://dx.doi.org/10.1103/PhysRevB.96.035151},
  doi     = {10.1103/PhysRevB.96.035151}
}

@article{Katsnelson2002,
  author  = {Katsnelson, M. I. and Lichtenstein, A. I.},
  title   = {Electronic structure and magnetic properties of correlated metals: A local self-consistent perturbation scheme},
  journal = {The European Physical Journal B},
  volume  = {30},
  pages   = {9--15},
  year    = {2002}, 
  url = {http://dx.doi.org/10.1140/epjb/e2002-00352-1},
  DOI = {10.1140/epjb/e2002-00352-1},
}

@article{Granas2012,
  author  = {Granas, O. and Di Marco, I. and Thunstrom, P. and Nordstrom, L. and Eriksson, O. and Bjorkman, T. and Wills, J. M.},
  title   = {Charge self-consistent dynamical mean-field theory based on the full-potential linear muffin-tin orbital method: Methodology and applications},
  journal = {Computational Materials Science},
  volume  = {55},
  pages   = {295--302},
  year    = {2012},
  url = {http://dx.doi.org/10.1016/j.commatsci.2011.11.032},
  DOI = {10.1016/j.commatsci.2011.11.032},
}

@article{Peters2014,
  author  = {Peters, L. and Di Marco, I. and Thunstrom, P. and Katsnelson, M. I. and Kirilyuk, A. and Eriksson, O.},
  title   = {Treatment of $4f$ states of the rare earths: The case study of {TbN}},
  journal = {Physical Review B},
  volume  = {89},
  pages   = {205109},
  year    = {2014},
  url = {http://dx.doi.org/10.1103/PhysRevB.89.205109},
  doi     = {10.1103/PhysRevB.89.205109}
}

@article{Zgid2012,
  author  = {Zgid, D. and Gull, E. and Chan, G. K.-L.},
  title   = {Truncated configuration interaction expansions as solvers for correlated quantum impurity models and dynamical mean-field theory},
  journal = {Physical Review B},
  volume  = {86},
  pages   = {165128},
  year    = {2012},
  url = {http://dx.doi.org/10.1103/PhysRevB.86.165128},
  doi     = {10.1103/PhysRevB.86.165128}
}

@article{Herzog2025,
  author  = {Herzog, B. and Thunstrom, P. and Eriksson, O.},
  title   = {A configuration interaction approach to solve the {Anderson} impurity model; applications to elemental {Ce}},
  journal = {npj Computational Materials},
  volume  = {11},
  pages   = {373},
  year    = {2025},
  url = {http://dx.doi.org/10.1038/s41524-025-01883-0},
  doi     = {10.1038/s41524-025-01883-0}
}

@article{magcal,
  title = {Magnetocaloric effect in {CeFe$_2$} and Ru-doped {CeFe$_2$} alloys},
  volume = {39},
  ISSN = {1361-6463},
  url = {http://dx.doi.org/10.1088/0022-3727/39/6/002},
  DOI = {10.1088/0022-3727/39/6/002},
  number = {6},
  journal = {Journal of Physics D: Applied Physics},
  publisher = {IOP Publishing},
  author = {Chattopadhyay,  M K and Manekar,  M A and Roy,  S B},
  year = {2006},
  month = Mar,
  pages = {1006–1011}
}

@article{szpunzar,
title = {Density of states and magnetic properties of {YCo$_5$} and {Y$_2$Co$_{17}$} compounds},
journal = {Physica B+C},
volume = {130},
number = {1},
pages = {29-33},
year = {1985},
issn = {0378-4363},
doi = {https://doi.org/10.1016/0378-4363(85)90174-3},
url = {https://www.sciencedirect.com/science/article/pii/0378436385901743},
author = {Barbara Szpunar}
}

@article{permmag,
  title = {Analysis of the influence mechanism of CeFe2 phase on coercivity in high-performance Nd-Ce-Fe-B diffusion magnets},
  volume = {623},
  ISSN = {0304-8853},
  url = {http://dx.doi.org/10.1016/j.jmmm.2025.173001},
  DOI = {10.1016/j.jmmm.2025.173001},
  journal = {Journal of Magnetism and Magnetic Materials},
  publisher = {Elsevier BV},
  author = {Qin,  Yuan and Wang,  Xiangming and Zhang,  Lele and Fan,  Min and Song,  Jie and Wang,  Gang and Liu,  Weiqiang and Chen,  Hao and Yue,  Penghao and Zhao,  Zhibo and Wang,  Zhanjia and Li,  Yuqing and Yue,  Ming},
  year = {2025},
  month = July,
  pages = {173001}
}

@article{heike,
  title = {Combining electronic structure and many-body theory with large databases: A method for predicting the nature of {$4f$} states in Ce compounds},
  volume = {1},
  ISSN = {2475-9953},
  url = {http://dx.doi.org/10.1103/PhysRevMaterials.1.033802},
  DOI = {10.1103/physrevmaterials.1.033802},
  number = {3},
  journal = {Physical Review Materials},
  publisher = {American Physical Society (APS)},
  author = {Herper,  H. C. and Ahmed,  T. and Wills,  J. M. and Di Marco,  I. and Bj\"{o}rkman,  T. and Iuşan,  D. and Balatsky,  A. V. and Eriksson,  O.},
  year = {2017},
  month = Aug 
}

@article{Brooks1989,
  title = {3d-5d band magnetism in rare earth transition metal intermetallics: {LuFe$_2$}},
  volume = {1},
  ISSN = {1361-648X},
  url = {http://dx.doi.org/10.1088/0953-8984/1/34/004},
  DOI = {10.1088/0953-8984/1/34/004},
  number = {34},
  journal = {Journal of Physics: Condensed Matter},
  publisher = {IOP Publishing},
  author = {Brooks,  M S S and Eriksson,  O and Johansson,  B},
  year = {1989},
  month = Aug,
  pages = {5861–5874}
}

@article{HIAsolver,
  title = {$\textit{Ab initio}$ calculations of quasiparticle band structure in correlated systems: {LDA}++ approach},
  volume = {57},
  ISSN = {1095-3795},
  url = {http://dx.doi.org/10.1103/PhysRevB.57.6884},
  DOI = {10.1103/physrevb.57.6884},
  number = {12},
  journal = {Physical Review B},
  publisher = {American Physical Society (APS)},
  author = {Lichtenstein,  A. I. and Katsnelson,  M. I.},
  year = {1998},
  month = Mar,
  pages = {6884–6895}
}

@misc{LSDAcheck, 
    note={This was tested in the present calculations; data not shown.}}

@Article{ce_alpha_1,
  title		= {The alpha-gamma transition of {Cerium} is entropy-driven},
  volume	= {96},
  doi		= {10.1103/PhysRevLett.96.066402},
  language	= {en},
  urldate	= {2025-02-20},
  journal	= {Physical Review Letters},
  author	= {Amadon, B. and Biermann, S. and Georges, A. and
		  Aryasetiawan, F.},
  year		= {2006},
  note		= {arXiv:cond-mat/0504732},
  pages		= {066402},
}

@article{ce_alpha_2,
  title = {Cerium Volume Collapse: Results from the Merger of Dynamical Mean-Field Theory and Local Density Approximation},
  volume = {87},
  ISSN = {1079-7114},
  url = {http://dx.doi.org/10.1103/PhysRevLett.87.276404},
  DOI = {10.1103/physrevlett.87.276404},
  number = {27},
  journal = {Physical Review Letters},
  publisher = {American Physical Society (APS)},
  author = {Held,  K. and McMahan,  A. K. and Scalettar,  R. T.},
  year = {2001},
  month = Dec 
}

@article{ce_alpha_3,
  title = {Thermodynamics of the {$\alpha - \gamma$} transition in cerium from first principles},
  volume = {89},
  ISSN = {1550-235X},
  url = {http://dx.doi.org/10.1103/PhysRevB.89.195132},
  DOI = {10.1103/physrevb.89.195132},
  number = {19},
  journal = {Physical Review B},
  publisher = {American Physical Society (APS)},
  author = {Bieder,  J. and Amadon,  B.},
  year = {2014},
  month = May 
}

\appendix 

\section{Details of calculations}

Calculations were performed with the full-potential linear muffin-tin orbital code, RSPt, in the local-density approximation (LDA) \cite{Granas2012}. 
The relativistic, non-spin-polarized LDA Hamiltonian was used as the starting point for the DMFT calculation. 
The experimental C15 structure was used, with a conventional cubic lattice constant of $a=7.304$ Å and a six-atom primitive cell containing two formula units.
The Brillouin zone was sampled using a $12\times12\times12$ $k$-point mesh.

For the DMFT calculations, the Fe $3d$, Ce $5d$, and Ce $4f$ muffin-tin heads were treated as separate correlated subspaces. 
The impurity problem of Fe $3d$ states was solved using spin-polarized T-matrix fluctuation exchange (SPTF) \cite{Katsnelson2002}, an approach that considers the influence of a local Hubbard U perturbatively.
For Ce $5d$ orbitals we used the static Hartree-Fock approximation (+U) and for the most correlated orbitals, the  
Ce $4f$s, we used either the Hubbard-I\cite{HIAsolver} or CLIC solver\cite{Zgid2012,Herzog2025} (The latter implementation can be found at \url{https://github.com/bslhrzg/clic}). 
The Hubbard-I approximation preserves the local Ce multiplet structure but ignores hybridization effects, which removes the hybridization bath from the impurity self-energy. 
In Hubbard-I, removal of the Ce-$4f$ hybridization bath prevents this magnetic environment from being transmitted directly to the atomic impurity. We therefore used the one-body $f$-$d$ exchange splitting \cite{Peters2014}
\begin{equation}
  \Delta_{fd}=I_{fd}m_d.
\end{equation}
We set $I_{fd}=0.01$~Ry, consistent in magnitude with the atomic Ce $4f-5d$ exchange integral  $J_{4f,5d} = 9$~mRy calculated by Brooks \textit{et al.}\cite{Brooks1989}. 

The SPTF+Hubbard-I calculation was converged charge self-consistently. 
In the subsequent CLIC calculation, the explicit spin-dependent hybridization function provided the polarization mechanism and the model $f$-$d$ exchange term was switched off. 

The bath-coupled, configuration-interaction (CI) treatment to the impurity problem retains local Coulomb interactions, spin-orbit coupling, and bath-induced charge fluctuations in the same many-body problem\cite{Zgid2012,Herzog2025}, which we refer to as the CLIC solver.

The fully localized limit double counting term was used for Fe $3d$ and Ce $5d$ orbitals, while the nominal-occupancy form was used for the Ce $4f$ states. 
All correlated calculations were performed at an electronic temperature of $0.001$~Ry.

A small symmetry-breaking field was used only to initialize the Fe $3d$ polarization and was set to zero immediately afterward. 
All subsequent iterations were performed without an external field, so the converged ordered state was sustained by the DMFT self-energies. 
This construction avoids combining a pre-existing LDA exchange splitting with the magnetic impurity self-energy; 
the double-counting correction therefore removes only the charge contribution already represented by LDA.

CLIC was applied in a single shot to the converged SPTF+Hubbard-I electronic structure. With CLIC, the impurity problem was solved using the previously described selected configuration interaction solver \cite{Herzog2025}. The diagonal components of the Matsubara-axis hybridization function were fitted using two bath levels per impurity spin-orbital, minimizing a weighted least-square cost to emphasize low frequencies; The original impurity-bath basis was used (so-called star geometry) with three configuration-selections iterations, starting from a reference state with one electron in the impurity. Thermally averaged Green's functions were evaluated using Lanczos recursion in particle addition and particle removal space expanded by one Hamiltonian application. 

\section{Magnetic moments table}
\begin{table}[H]
\centering
  \caption{Magnetic moments in CeFe$_2$, in $\mu_B$. Fe values are per atom and the final column is per formula unit. Positive values are parallel to the Fe magnetization. Theoretical local values are muffin-tin-head moments. The LSDA formula-unit result (superscript $a$) is obtained from the full cell, including the interstitial contribution. The DMFT formula-unit values (superscript $b$) are sums of the displayed MT-head contributions and constitute the complete magnetization within that representation.}
  \label{tab:moments}
  \setlength{\tabcolsep}{5.0pt}
\begin{adjustbox}{max width=\columnwidth}
  \begin{tabular}{lrrrrrrrr}
    \toprule
    & \multicolumn{3}{c}{Fe} & \multicolumn{1}{c}{Ce $5d$} & \multicolumn{3}{c}{Ce $4f$} & \multicolumn{1}{c}{CeFe$_2$} \\
    \cmidrule(lr){2-4}\cmidrule(lr){5-5}\cmidrule(lr){6-8}\cmidrule(lr){9-9}
    Method & $m_L$ & $m_S$ & $m$ & $m$ & $m_L$ & $m_S$ & $m$ & $m$ \\
    \midrule
    Spin-polarized LSDA ref. & 0.05 & 1.66 & 1.71 & -0.21 & 0.18 & -0.49 & -0.31 & $2.70^{a}$ \\
    SPTF+HIA & 0.18 & 1.31 & 1.49 & -0.09 & 0.43 & -0.14 & 0.29 & $3.18^{b}$ \\
    SPTF+CLIC & 0.17 & 1.11 & 1.28 & -0.08 & 0.28 & -0.34 & -0.06 & $2.42^{b}$ \\
    \midrule
    XMCD \cite{Delobbe1998} & 0.10 & 1.24 & 1.34 & -0.13 & 0.21 & -0.37 & -0.16 & 2.39 \\
    Neutron \cite{Kennedy1993} & -- & -- & 1.17 & -0.07 & 0.03 & -0.10 & -0.07 & 2.20 \\
    \bottomrule
  \end{tabular}
  \end{adjustbox}

\end{table}

\end{document}